\documentclass[aps, twocolumn, letter, showpacs, floatfix,longbibliography]{revtex4-2}

\usepackage{amsmath, amssymb, physics}
\usepackage[hypertexnames=false]{hyperref}
\usepackage{datetime}
\usepackage{enumitem}
\usepackage{graphicx}
\usepackage{esdiff}
\usepackage{MnSymbol}
\usepackage{siunitx}
\usepackage{circuitikz}
\usepackage[normalem]{ulem}
\renewcommand{\selectlanguage}[1]{}
\hypersetup{
    colorlinks,
    linkcolor={red!50!black},
    citecolor={blue!50!black},
    urlcolor={blue!80!black}
}

\def\kup{\left\vert\uparrow\right\rangle}
\def\kdn{\left\vert\downarrow\right\rangle}

\footnotetext{These authors contributed equally to this work}

\begin{document}

\makeatletter
\def\@fnsymbol#1{\ensuremath{\ifcase#1\or *\or \ddagger\or \mathsection\or \mathparagraph\or \|\or **\or \dagger\dagger
   \or \ddagger\ddagger \else\@ctrerr\fi}}
\makeatother

\title{
Saturation of thermal and spin conductances in a dissipative superfluid junction
}

\author{Meng-Zi Huang\textsuperscript{\dagger}}
\email{mhuang@phys.ethz.ch}
\author{Philipp Fabritius\textsuperscript{\dagger}}
\altaffiliation{Present address: X-Rite Europe GmbH, Regensdorf, Switzerland}
\author{Jeffrey Mohan}
\altaffiliation[Present address: ]{Welinq SAS, Paris, France}
\author{Mohsen Talebi}
\author{Simon Wili}
\author{Tilman Esslinger}
\affiliation{Institute for Quantum Electronics and Quantum Center, ETH Zurich, 8093 Zurich, Switzerland}

\date{2025-06-16}

\begin{abstract}
Fermionic superfluid junctions typically exhibit suppressed thermal and spin transport due to the presence of a pairing gap but allow coherent particle transport. While dissipation generally weakens coherent transport, it can also induce excitations that open other transport channels. In this work, we experimentally study a one-dimensional superfluid junction of strongly interacting fermions with local particle loss and observe dissipation-induced thermal and spin transport that appear to saturate at strong dissipation. 
Notably, in this regime, the measured thermal and spin conductances are comparable to the universal quantized conductance of one-dimensional ideal Fermi gas.
Qualitatively similar behavior is observed
for two dissipation mechanisms, either spin-imbalanced or pairwise losses. Our findings provide new insights into transport in interacting open quantum systems and suggest possibilities of dissipative control of spin and thermoelectric transport.
\end{abstract}

\maketitle

\emph{Introduction---}
Quantum gases have greatly contributed to the understanding of strongly interacting matter both at equilibrium~\cite{zwierlein_thermodynamics_2016,ku_revealing_2012, li_observation_2024} and out of equilibrium~\cite{eisert_quantum_2015}. In particular, fermionic lithium, with its broad Feshbach resonances, offers a unique model system featuring, for example, robust superfluidity with strong interactions~\cite{zwerger_bcs-bec_2012}.
More recently, highly controllable quantum gas experiments also provide test beds for physics of open quantum systems, where many-body coherence competes with dissipation~\cite{labouvie_bistability_2016,tomita_observation_2017,dogra_dissipation-induced_2019,bouganne_anomalous_2020,dreon_self-oscillating_2022}. Challenges for studying strongly interacting systems under dissipation, such as detrimental heating, can be alleviated by transport settings with localized dissipation~\cite{ barontini_controlling_2013,labouvie_bistability_2016,corman_quantized_2019,huang_superfluid_2023,gievers_quantum_2024}. However, experiments beyond particle transport, although having been studied in closed systems~\cite{yan_thermography_2024,li_second_2022,luciuk_observation_2017,sommer_universal_2011, krinner_mapping_2016, valtolina_exploring_2017, dogra_universal_2023,fabritius_irreversible_2024}, are still rare in open systems with controlled dissipation. For example, one-dimensional (1D) junctions between two fermionic superfluids exhibit enhanced particle transport~\cite{leggett_quantum_2006, husmann_connecting_2015} but suppressed thermal and spin transport compared to Fermi liquids due to pairing in the reservoirs~\cite{husmann_breakdown_2018, pershoguba_thermopower_2019, krinner_mapping_2016,sekino_mesoscopic_2020, fabritius_irreversible_2024} . With particle dissipation in the junction, coherent transport is weakened by the suppression of pairing~\cite{huang_superfluid_2023, visuri_dc_2023}, which can instead facilitate thermal and spin transport. Yet, such scenarios have been so far unexplored both theoretically and experimentally.

Here, we combine local dissipation with two-terminal measurements of particle, entropy, and spin currents, revealing an unexpected state of the strongly interacting system with saturating thermal and spin conductances. In our 1D ballistic junction connecting two unitary Fermi gases in the superfluid regime, the particle transport displays a nonlinear current-bias relation~\cite{husmann_connecting_2015}, bearing similarities to transport in superconducting quantum point contacts~\cite{cuevas_hamiltonian_1996, scheer_conduction_1997}. However, the conventional picture from superconducting systems~\cite{husmann_connecting_2015} is confronted by recent observation of an abnormally large Seebeck coefficient, i.e., a large \emph{advective} entropy current carried by the particle current~\cite{fabritius_irreversible_2024}. 
In contrast, thermal conductance characterizes entropy \emph{diffusion}, which can be discriminated from the advective current given their distinct timescales \cite{fabritius_irreversible_2024}.
As a result, thermal conductance can be obtained from a thermoelectric model~\cite{grenier_thermoelectric_2016} generalized to nonlinear response~\cite{fabritius_irreversible_2024}.

In the present work, we engineer particle dissipation in the superfluid junction and observe a sharp increase in diffusive thermal and spin transport. Interestingly, both the thermal and spin conductances seem to approach the quantum-limited values for a non-interacting system without dissipation~\cite{pendry_quantum_1983, blencowe_universal_2000, jezouin_quantum_2013, cui_quantized_2017,krinner_observation_2014}.
Moreover, qualitatively similar saturating behavior is observed for two different
dissipation mechanisms in regard to pairing---a spin-imbalanced loss that dissipates one spin more than the other, leaving some broken pairs; and pairwise losses that are strictly spin correlated. 

\begin{figure}
    \centering
    \includegraphics[width=85mm]{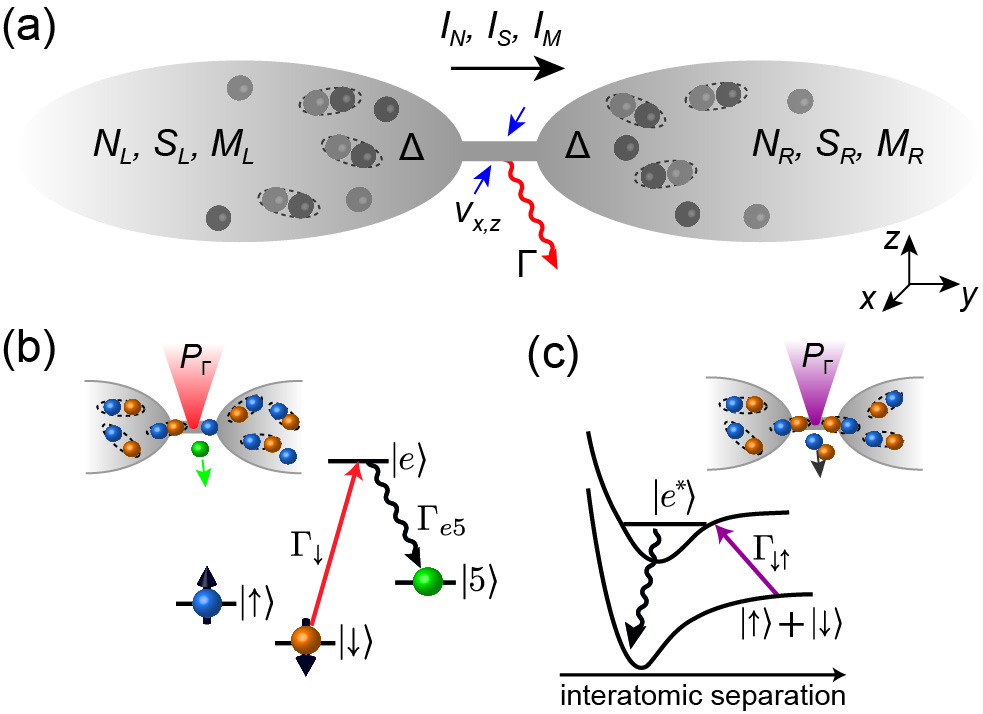}
    \caption{\textbf{Scheme to measure transport coefficients of a superfluid junction in the presence of engineered spin-imbalanced or pairwise dissipation.} (a) Measuring entropy $I_S$, particle $I_N$ and spin $I_M$ currents probes the dissipative, quasi-1D channel (transversal confinement $h\nu_{x,z}\gtrsim 5k_B T$) connecting two fermionic superfluid reservoirs ($L$ and $R$) with pairing gap $h\Delta\gg k_B T$ near the entrance to the channel. The local particle loss rate $\Gamma$ is engineered via spin-imbalanced (b) and pairwise (c) dissipation, realized by a focused laser beam inside the channel with a waist smaller than the channel length. (b),(c) Illustration and energy diagram of the optical pumping process (b) with rate $\Gamma_\downarrow$ to produce spin-imbalanced particle losses and the photoassociation of molecules at rate $\Gamma_{\uparrow\downarrow}$ (c) to realize strictly pairwise losses.  
    }\label{fig:fig1}
\end{figure}

\emph{Experiment and dissipation mechanisms---}
We start all measurements by preparing a degenerate Fermi gas of $^6$Li in a balanced mixture of the lowest ($\kdn$) and
third-lowest ($\kup$) hyperfine ground states in the unitary regime (Feshbach resonance at 689 G) with an initial total atom number $N=N_\downarrow+N_\uparrow=2.9(2)\times10^5$. The gas is trapped in a near-harmonic trap separated into two halves along the $y$ axis by two intersecting repulsive TEM$_{01}$-like beams. This forms two reservoirs, left ($L$) and right ($R$), connected by a ballistic channel.
The transverse confinement of the channel is in the quasi-1D limit with trapping frequencies $\nu_z=\SI{9.4(1)}{kHz}$ along $z$ and $\nu_x=\SI{10.9(2)}{kHz}$ along $x$, such that $h\nu_{x,z}\gtrsim 5k_B T$, where $h$ is Planck's constant, $k_B$ is the Boltzmann constant and $T$ is the temperature~\cite{fabritius_irreversible_2024, husmann_connecting_2015}. 
Transport between the reservoirs can be initiated by preparing different types of imbalances between the reservoirs \cite{krinner_two-terminal_2017}.
An attractive ``gate'' beam covering the channel region is used to adjust the local chemical potential during transport.
After transport between the reservoirs for a time $t$, we perform thermometry of each reservoir in a half harmonic trap by adiabatically removing the channel confining beams while keeping the two reservoirs separated by a thin ``wall'' beam~\cite{fabritius_irreversible_2024}. As the entropy $S_i=S_{i,\uparrow}+S_{i,\downarrow}$ and atom number $N_i=N_{i,\uparrow}+N_{i,\downarrow}$ in each reservoir ($i=L,R$) are conserved between the end of transport and imaging, we use these two thermodynamic quantities to describe the reservoirs throughout the paper. 
The prepared reservoirs typically have an initial average entropy of $s=S/N k_B=1.58(6)$ which is below the superfluid transition at $s_c=1.9(1)$
in the slightly anharmonic trap during transport~\cite{fabritius_irreversible_2024}. 
For a first set of measurements, an entropy imbalance ($\Delta S=S_L - S_R\approx 0.15 N k_B$) between the reservoirs is prepared while keeping the atom number imbalance negligible, $|\Delta N/N|<4\%$ where $\Delta N=N_L-N_R$. This initial state allows us to measure the advective and diffusive response of the system~\cite{fabritius_irreversible_2024, husmann_breakdown_2018}. For a second set of measurements, we prepare a pure magnetization imbalance $\Delta M = M_L-M_R\approx 0.13N$, where $M_i = N_{i,\uparrow}-N_{i,\downarrow}$, with negligible atom number and entropy imbalances  $|\Delta N/N|<2\%$, $|\Delta S/Nk_B| < 5\%$. This preparation allows us to probe the spin diffusion under the same conditions of the channel~\cite{krinner_mapping_2016, fabritius_irreversible_2024}.

To realize controlled particle dissipation inside the channel, a resonant laser beam propagating along $z$ is tightly focused into the channel with a $1/e^2$ beam radius of $w_\Gamma=\SI{1.3(1)}{\micro\meter}$, similar to the channel width and shorter than the channel length. 
Two types of dissipation are achieved with this beam by tuning its frequency. First, spin-imbalanced particle loss is engineered by a resonant excitation of $\ket{\downarrow}$, optically pumping the atoms to an auxiliary ground state that leaves the system [Fig.~\ref{fig:fig1}(b)] \cite{huang_superfluid_2023}
with a single-atom photon scattering rate $\Gamma_\downarrow$. Because of attractive interactions, atoms of the other spin $\ket{\uparrow}$ are also lost, though at a lower rate of $0.83(4)\times\Gamma_\downarrow$. Second, pairwise loss is engineered via a photoassociation transition that excites a pair of $\ket{\uparrow}$ and $\ket{\downarrow}$ to an excited molecular state~\cite{partridge_molecular_2005,werner_number_2009,wang_photoexcitation_2021} which subsequently leaves the trap [Fig.~\ref{fig:fig1}(c)] 
with a photon-scattering rate $\Gamma_{\uparrow\downarrow}$. The loss rate $\Gamma_{\uparrow\downarrow}$ depends on both the atomic density and the pair correlations and thus on Tan's contact \cite{jager_precise_2024}. 
The power of the dissipative beam is varied up to $P_\mathrm{max} = \SI{116(2)}{pW}$ corresponding to an intensity $I_\mathrm{max}=\SI{44(2)}{W/m^2}$ at the center of the beam, and a maximal $\Gamma_\downarrow=\SI{93(2)}{ms^{-1}}$ for spin-imbalanced dissipation. We rely on the measured atom loss for comparison between $\Gamma_{\uparrow\downarrow}$ and $\Gamma_\downarrow$. 
Because of the locality of the dissipation, neither method induces substantial heating to the reservoirs~\cite{supplement}. 
\nocite{fabritius_irreversible_2024,partridge_molecular_2005,wang_photoexcitation_2021,ku_revealing_2012,husmann_breakdown_2018,krinner_mapping_2016,grenier_thermoelectric_2016}

\begin{figure}[t]
    \centering
    \includegraphics[width=85mm]{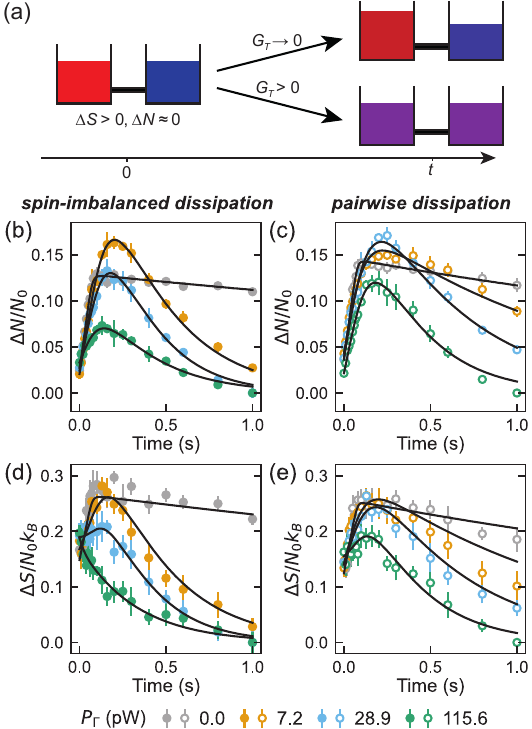}
    \caption{\textbf{Coupled particle and thermal transport revealing entropy diffusion enhanced by both spin-imbalanced and pairwise dissipation.} System evolution in normalized particle (b),(c) and entropy (d),(e) imbalances starting from a pure entropy imbalance [illustrated in (a)], with spin-imbalanced (b),(d) and pairwise dissipation (c),(e). The initial advective response is enhanced by weak dissipation [e.g., orange circles in (b)]. The dissipation strength is quantified by the optical power of the dissipation beam, $P_\Gamma$. The diffusion timescale after the initial response decreases monotonically with increasing $P_\Gamma$. 
    Markers (error bars) correspond to averages (standard deviations) over $5$ repetitions. Solid curves are fit to the phenomenological model (see text). }\label{fig:fig2}
\end{figure}

\emph{Thermal transport measurements---}In the first set of measurements we focus on coupled entropy and particle transport, by starting the experiment with no particle imbalance $\Delta N\approx0$ but an entropy imbalance $\Delta S>0$ [Fig.~\ref{fig:fig2}(a)]. The system dynamics are analogous to the Seebeck effect in thermoelectric materials where a current is induced by a temperature difference. The advective current in our system dominates the initial response. It is driven by both the chemical potential bias $\Delta\mu$ and temperature bias $\Delta T$, with a total thermodynamic force $\Delta\mu + \alpha_c\Delta T$, where $\alpha_c$ is the Seebeck coefficient, i.e., the transported entropy per particle. After this driving force is balanced with a remaining temperature imbalance $\Delta T>0$, thermal diffusion provides a way for the system to fully relax to equilibrium [Fig.~\ref{fig:fig2}(a)]. The timescale of the diffusive relaxation is characterized by the 
thermal conductance $G_T$
as the diffusive entropy current responds linearly to $\Delta T$. 

In the absence of dissipation, 
the particle- and entropy-current response to $\Delta\mu$ and $\Delta T$ is highly nonlinear as observed previously \cite{husmann_connecting_2015,fabritius_irreversible_2024}. This leads to a highly non-exponential rise in both $\Delta N$ and $\Delta S$ [normalized by $N(t=0)\equiv N_0$, gray circles in Figs.~\ref{fig:fig2}(b)--(e)] and a sudden halt, when the advective response is balanced, followed by a very slow decay. 
The final state resembles a non-equilibrium steady state~\cite{fabritius_irreversible_2024,husmann_breakdown_2018} in which the thermal diffusion is negligible.
In contrast, by applying either type of dissipation [in the order of increasing strength: orange, blue, and green circles in Figs.~\ref{fig:fig2}(b)--(e)], the initial response is slightly slower but the following timescale of diffusion becomes much faster, almost relaxing the system to equilibrium within $\SI{1}{s}$. Notably, dissipation qualitatively reduces the nonlinear nature of the advective response leading to a smoother evolution of $\Delta N$ and $\Delta S$. 

In order to quantify the changes in advection and diffusion with dissipation, we fit the phenomenological model developed in Ref.~\cite{fabritius_irreversible_2024} for closed systems, which is a conventional thermoelectric model generalized to nonlinear response~\cite{grenier_thermoelectric_2016, pavelka_multiscale_2018}. 
The model consists of a \emph{nonlinear} particle current $I_N = I_\mathrm{exc} \tanh\pqty{\frac{\Delta\mu + \alpha_c \Delta T}{\sigma}}$ and an entropy current containing an advective part proportional to $I_N$ and a linear diffusion: $I_S = \alpha_c I_N + G_T \Delta T/T$.
$I_\mathrm{exc}$ is the ``excess'' current the system reaches after saturating the nonlinearity (biases exceeding $\sigma$)~\cite{fabritius_irreversible_2024,cuevas_hamiltonian_1996} and $\sigma$ is an energy scale that determines the nonlinearity of the current response.
Together with the linear reservoir response~\cite{supplement, grenier_thermoelectric_2016}, 
the dynamics of the reservoirs in the $(\Delta N, \Delta S)$ space can be fully described. 
Although the phenomenological model does not explicitly include losses, which would otherwise involve additional assumptions and fit parameters~\cite{uchino_comparative_2022,roberts_hidden_2021}, we find that the model can well fit the data with dissipation when $I_N$ and $I_S$ represent the \emph{apparent} currents: $I_N =(-1/2)\dv*{\Delta N}{t}$ and $I_S =(-1/2)\dv*{\Delta S}{t}$. 
The effect of the dissipation is encapsulated in changes of the transport coefficients such as $I_\mathrm{exc}$, $\sigma$, $\alpha_c$, and $G_T$. 
The fits for varying dissipation strength are shown as solid curves in Figs.~\ref{fig:fig2}(b)--(e). They all share the same reservoir response coefficients so the fitted transport coefficients can reveal qualitative changes of the two transport modes. We will discuss these results further below in Fig.~\ref{fig:fig4}.
Notably, the qualitative behavior of $G_T$, which is the main focus of this work, can be obtained from the exponential timescale of the slow relaxation in $\Delta N$ without assuming the nonlinear phenomenological model~\cite{supplement}. 

The enhancement of the ``Seebeck'' response---maximal $\Delta N$ reached---for weak dissipation strength [orange data in Fig.~2(b), orange and blue data in Fig.~2(c)]  
remains to be fully understood microscopically.
Phenomenologically, as in conventional thermoelectric models for finite systems \cite{grenier_thermoelectric_2016}, the response in $\Delta N$ induced by $\Delta S$ is determined by $\alpha_r-\alpha_c$, i.e., a balance between the channel Seebeck property $I_N\sim \alpha_c\Delta T$ and the reservoir response $\Delta\mu\sim-\alpha_r\Delta S$, where $\alpha_r=(\partial S/\partial N)_T$ \cite{grenier_thermoelectric_2016}. The observation can result from a 10\% increase of $\alpha_r-\alpha_c$ by dissipation as shown in the fitted results~\cite{supplement}. Although we only fit $\alpha_c$ to limit the degrees of freedom, both $\alpha_c$ and $\alpha_r$ could be affected by dissipation. A possible scenario is an increase of $\alpha_r$ compared to $\alpha_c$ due to residual heating or dissipation preferentially removing low energy atoms, reducing the local degeneracy ($\alpha_r$ increases with decreasing degeneracy).  

\begin{figure}[t]
    \centering
    \includegraphics[width=85mm]{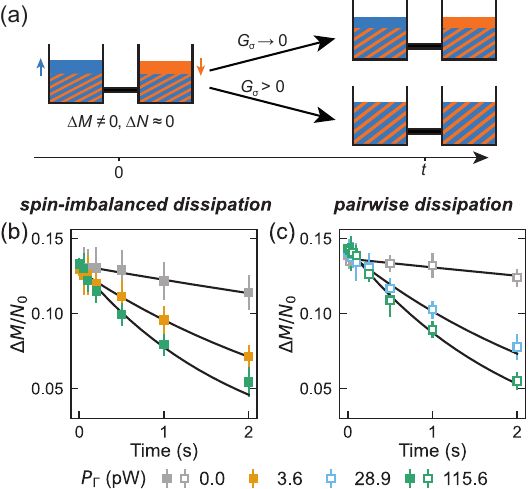}
    \caption{\textbf{Spin diffusion between superfluid reservoirs enhanced by both spin-imbalanced and pairwise dissipation.} (a) Sketch of the initial $\Delta M\neq0,\,\Delta N\approx0$, and possible final states of the system with and without spin diffusion. (b) The relative magnetization $\Delta M/N_0$, plotted as a function of time, relaxes faster with increasing spin-imbalanced dissipation (from gray to green). (c) Applying pairwise dissipation in the quasi-1D junction also leads to decay of the relative magnetization. Solid black lines are fits to exponential decays. Markers (error bars) correspond to averages (standard deviations) over $5$ repetitions.
    }\label{fig:fig3}
\end{figure}

\emph{Spin transport measurements---}In the second set of measurements, we explore the effect of dissipation on spin transport through the junction.
In contrast to the advective entropy current, the spin current $I_M$ responds linearly to an applied spin bias $\Delta b=(\Delta\mu_\uparrow-\Delta\mu_\downarrow)/2$, such that 
{$I_M=G_\sigma\Delta b$ where $G_\sigma$ is the spin conductance.}
Figure~\ref{fig:fig3}(a) illustrates the initial condition and possible final conditions in the system with different $G_\sigma$. The initial state is prepared to have a pure magnetization imbalance $\Delta M_0 = M_L-M_R\neq0$, ($M_i = N_{i,\uparrow}-N_{i,\downarrow}$) and almost no initial particle or entropy imbalance $\Delta N_0 \approx \Delta S_0 \approx 0$ \cite{supplement}. The subsequent relative magnetization $\Delta M(t)/N_0$ dynamics are displayed in Figs.~\ref{fig:fig3}(b),(c) for a selection of measured dissipation strengths. Without dissipation (gray squares) the magnetization imbalance hardly relaxes within the explored timescale~(2\,s), indicating spin-insulating behavior in line with previous results~\cite{krinner_mapping_2016}. Spin-imbalanced dissipation (b) and pairwise dissipation (c) inside the junction both lead to a faster spin relaxation, approaching equilibrium within the experimental timescale. In order to quantify the change of spin diffusion with dissipation, an offset-free exponential decay is fitted [solid lines in Figs.~\ref{fig:fig3}(b),(c)]. The spin 
{conductance $G_\sigma$}
is calculated from the fitted decay time constant~\cite{krinner_mapping_2016, supplement}.

\begin{figure}[t]
    \centering
    \includegraphics[width=85mm]{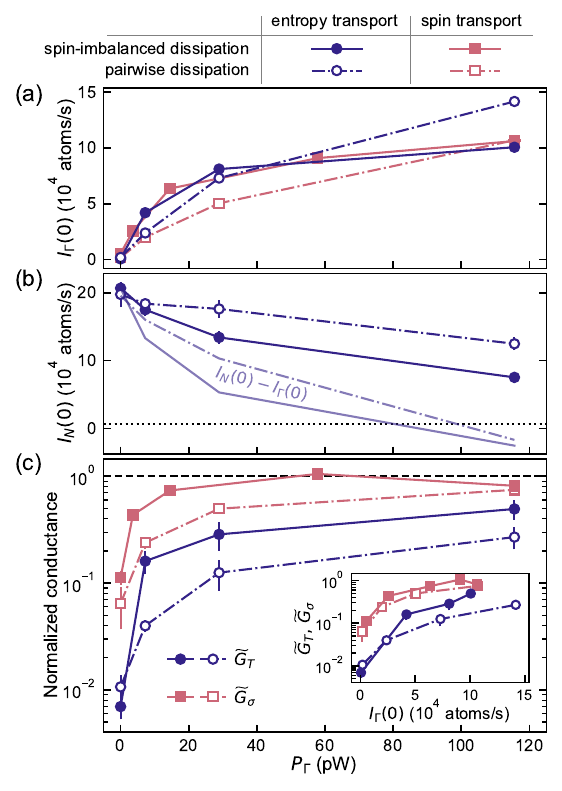}
    \caption{\textbf{Extracting thermal and spin conductances versus dissipation strength, showing that they approach values for non-interacting fermions.} (a) The average loss current $I_\Gamma(0)$ from each reservoir is shown as a function of the dissipation power $P_\Gamma$. Navy circles (pink squares) correspond to the entropy (spin) transport experiment, and filled (open) symbols correspond to spin-imbalanced (pairwise) dissipation. Transport coefficients are obtained from fits to the dynamics observed in the entropy transport (circles) and spin transport (squares) measurements, displayed in Fig.~\ref{fig:fig2} and Fig.~\ref{fig:fig3}, respectively. (b) Initial particle current $I_N(0)$.
    The dotted horizontal line indicates the expected particle current in a non-interacting, non-dissipative ballistic 1D junction. The lines in lighter color are the lower bounds for the conserved current given by $I_N(0)-I_\Gamma(0)$ (see text). (c) The normalized spin conductance $\widetilde{G}_\sigma$ (squares) and thermal conductance $\widetilde{G}_T$ (circles) are plotted versus the dissipation strength and versus the initial loss current (inset). The unity value (dashed horizontal line) corresponds to a non-interacting ballistic 1D junction without dissipation.
    Error bars correspond to standard errors from the fit. 
    }
    \label{fig:fig4}
\end{figure}

\emph{Quantifying the saturation of thermal and spin conductances---}We now examine how the entropy and spin diffusion scales with the dissipation strength. 
The fitted transport characteristics versus dissipation strength are summarized in Fig.~\ref{fig:fig4}.
Figure~\ref{fig:fig4}(a) shows the initial loss current $I_\Gamma(0)=\gamma_N N_0/2$ from all experiments, 
obtained from an exponential fit to the total atom number 
 $N(t)=N_0e^{-\gamma_N t}$. The loss current shows a trend of saturation as a function of the dissipation power $P_\Gamma$. Spin-imbalanced dissipation leads to higher losses at low $P_\Gamma$ and saturates faster compared to pairwise dissipation. 
Figure~\ref{fig:fig4}(b) shows the initial particle current $I_N(0)$ from the entropy transport experiment (Fig.~\ref{fig:fig2}) obtained from a linear fit to the initial evolution of $\Delta N$~\cite{supplement}. 
Similar to previous results~\cite{huang_superfluid_2023}, we find that dissipation decreases the particle current, but it remains much higher than 
the normal current expected in a non-interacting 1D junction $I_{\mathrm{n}0}=2n_\mathrm{m}\Delta\mu/h$ (dotted horizontal line), where $n_\mathrm{m}$ is the number of occupied transverse modes in the channel.
The observed current $I_N=-(\dot{N}_L-\dot{N}_R)/2$ can, in principle, arise purely from asymmetric losses, so we provide a lower bound for the \emph{conserved} current $I^\mathrm{cons}$---atoms transported through the junction without being lost. By writing $\dot{N}_{L/R} = \mp I^\mathrm{cons}-I^\mathrm{loss}_{L/R}$, where $I^\mathrm{loss}_{L/R}$ is the loss current from each reservoir, we have $I_N = I^\mathrm{cons} + (I^\mathrm{loss}_{L}-I^\mathrm{loss}_{R})/2$ and $I_\Gamma=(I^\mathrm{loss}_L+I^\mathrm{loss}_R)/2$. The lower bound for $I^\mathrm{cons}(0)$ is given by $I_N(0)-I_\Gamma(0)$, assuming the unlikely scenario where the lost atoms all come from one reservoir, $I^\mathrm{loss}_R=0$. These bounds are shown in the same line shape as $I_N(0)$ in lighter colors.

Figure~\ref{fig:fig4}(c) shows the normalized thermal 
conductance $\widetilde{G}_T$
(circles) as a function of the dissipation power $P_\Gamma$. They are obtained from fitting the phenomenological model to data from the entropy transport experiment (Fig.~\ref{fig:fig2})~\cite{supplement,fabritius_irreversible_2024}. Also plotted are the normalized spin 
conductance $\widetilde{G}_\sigma$
(squares, from Fig.~\ref{fig:fig3}) from the spin transport experiment. Both conductances are normalized by their non-interacting expectation in a non-dissipative 1D junction, i.e., 
$G_{T,0} = 2n_\mathrm{m}\pi^2 k_B^2 T/3h$ and $G_{\sigma,0}=2n_\mathrm{m}/h$, such that $\widetilde{G}_T,\,\widetilde{G}_\sigma=1$ 
correspond to the non-interacting values (dashed horizontal line). The factor of $2$ accounts for the two spins. 
The data points at $P_\Gamma=0$ should ideally be identical for different dissipation types (solid and open symbols). The discrepancies are likely due to small uncertainties and drifts in the alignment of the experiment, as well as fluctuations in the prepared atom number and temperature. 
We find that both dissipation types induce a sharp increase of more than 1 order of magnitude in spin conductance and almost 2 orders of magnitude in thermal conductance. Interestingly, both conductances seem to approach their values in the non-interacting regime without dissipation, which is a quantum limit given by the quantization of transverse modes in the 1D channel~\cite{pendry_quantum_1983, blencowe_universal_2000, jezouin_quantum_2013,cui_quantized_2017,krinner_observation_2014}. 
These conductances are much higher than the apparent transport purely arising from atom loss~\cite{supplement}, so the enhanced diffusion cannot be attributed to artifacts of losses. 
Notably, without using the phenomenological model, the saturating behavior of $G_T$ can also be obtained from the timescale of slow relaxation in $\Delta N$ (Fig.~\ref{fig:fig2}) by a simple exponential fit. An upper bound of $G_T$ can be estimated, which is also comparable to $G_{T,0}$~\cite{supplement}.
The same experiments have been performed with different transverse confinements $\nu_x$, hence, different $n_\mathrm{m}$ (within the quasi-1D regime), and the data of $G_T$ collapse when normalized by $n_\mathrm{m}$~\cite{supplement}. 

We can further compare the two types of dissipation by replotting the conductances versus the loss current [inset of Fig.~\ref{fig:fig4}(c)]. In this regard, the two mechanisms lead to similar spin and thermal diffusion for a given particle loss rate, with the pairwise dissipation being slightly less ``destructive'' to the superfluid junction. 
This difference, albeit small, agrees with the weaker suppression of particle current by pairwise loss [Fig.~\ref{fig:fig4}(b)], and could have the following origin. While the pairwise dissipation reduces the pairing field which could be replenished by the superfluid reservoirs, the spin-imbalanced dissipation creates unpaired quasi-particle excitations above the superfluid gap and is harder to remove.
At strong dissipation, both types of dissipation approach comparable limits. 

Although the microscopic mechanism of this saturating transport remains unclear, a possible picture is that thermal and spin conductances are initially suppressed by pairing, and breaking pairs by dissipation can recover both conductances. Notably, for a non-interacting dissipative system~\cite{corman_quantized_2019}, a similar dissipation strength
would strongly reduce the conductance $< 10^{-1}/h$, i.e., well below the universal quantized conductance. Our
data for the strongly interacting case suggest that the thermal and spin conductances approach the universal
values for quantized conductance, $1/h$ per mode.

\emph{Conclusions---}We have studied the effects of spin-imbalanced and pairwise dissipation on thermal and spin transport between two superfluids connected by a quasi-1D ballistic junction. We have found that dissipation leads to a remarkable increase in thermal and spin diffusion, and the conductances approach the quantum-limited, non-interacting values. 
Our observations call for a microscopic description of the system, for example, by extending the frameworks of the Lindblad-Keldysh field theory~\cite{sieberer_universality_2023} or the generalized hydrodynamics~\cite{doyon_generalized_2025}, that might be crucial to understanding the nature of the quantum-limited transport in the strongly interacting and strongly dissipative system. Future experiments exploring different types of excitation
to suppress pairing or going beyond 1D could further test the connection with the universal conductance limit. 
Moreover, the sharp increase of thermal and spin conductances at weak dissipation as well as the enhanced Seebeck response demonstrate a possibility to dissipatively control the ``thermoelectric'' and spin transport in a superfluid junction. These mechanisms may shed light on the design of functional thermoelectric or spintronic devices.

\emph{Acknowledgments---}
We thank Alexander Frank for his contributions to the electronics of the experiment. We are thankful for inspiring discussions with Sebastian Diehl, Alex G\'omez Salvador, Yi-Fan Qu, and Eugene Demler. We acknowledge the Swiss National Science Foundation (Grants No.~212168, UeM019-5.1, and TMAG-2\_209376) and European Research Council advanced grant TransQ (Grant No.~742579) for funding.

\bibliography{bibliography,supplement}

\clearpage
\section*{Supplemental Materials}
\setcounter{figure}{0}
\setcounter{table}{0}
\setcounter{section}{0}
\renewcommand{\thesection}{S\arabic{section}}
\renewcommand{\thefigure}{S\arabic{figure}}
\renewcommand{\thetable}{S\arabic{table}}
\renewcommand{\theequation}{S\arabic{equation}}

\section{Experimental details}
\subsection{Preparation}

The transport geometry is created by intersecting two repulsive TEM$_{01}$ like laser beams (along $x$ and $z$) at the center of a cigar-shaped harmonic trap, separating it into two half-harmonic reservoirs connected by a channel. The harmonic trapping frequencies are $\nu_x=\SI{171(1)}{Hz}$, $\nu_y=\SI{28.3(1)}{Hz}$ and $\nu_z=\SI{164(1)}{Hz}$. The channel forming beam along $x$ (``lightsheet'') has a waist of $w_{y,\mathrm{LS}}\approx\SI{30.2}{\micro\meter}$ along the transport direction $y$, forming a 2D region in the $x-y$ plane. Within this 2D region, the waist of the other channel forming beam along $z$ determines the length of the 1D channel, $w_{y,\mathrm{ch}}\approx\SI{6.8}{\micro\meter}$
(here the beam waists $w$ are defined as the $1/e^2$ beam radius). An attractive, circular ($w_{g}\approx\SI{31.8}{\micro\meter}$) ``gate'' beam, propagating along $z$, is centered on the channel to control the local potential $V_g/k_B= \SI{2.17(10)}{\micro K}$ at the channel contacts. A fast CCD camera is used to sequentially image the atomic clouds of each spin state after every experimental run. The first absorption image is taken of state $\ket{\downarrow}$ which leads to off-resonant photon scattering of state $\ket{\uparrow}$ that perturbs the shape of the cloud. This results in a slight overestimation $\sim 5\%$ of the entropy of $\ket{\uparrow}$ but we do not apply any correction for simplicity and being conservative. More details on the preparation and thermometry of the system can be found in our previous work~\cite{fabritius_irreversible_2024}. 

\subsection{Spin-imbalanced and pairwise dissipation}
A laser beam resonant at $\ket{\downarrow}\equiv\ket{2S_{1/2},m_J=-1/2,m_I=1}\rightarrow\ket{2P_{3/2},m_J=3/2,m_I=0}$ transition, with a waist of $w=\SI{1.3(1)}{\micro\meter}$ is projected onto the channel (along $z$) using a digital micro-mirror device for aberration correction. The used transition optically pumps atoms from the lowest hyperfine-state $\ket{\downarrow}$ to an auxiliary hyperfine state $\ket{5}\equiv\ket{2S_{1/2},m_J=1/2,m_I=0}$. This state is neither trapped, nor does it interact with  $\ket{\downarrow}$ or $\ket{\uparrow}\equiv\ket{2S_{1/2},m_J=-1/2,m_I=-1}$, such that atoms which scattered a photon leave the trap unimpededly. We calibrate the power and the intensity of the beam using both a powermeter and a camera. Using the calibration we can calculate the atom-photon scattering rate which is $\Gamma_\downarrow/P_\Gamma=\SI{0.80(1)}{ms^{-1}/pW}$. Generally, the intensities used in this paper are far below the saturation intensity of the transition $I_\mathrm{sat}=\SI{8.7}{kW/m^2}$.

Pairwise dissipation is engineered using the same beam, by tuning its frequency to a molecular photoassociation resonance (similar to Ref.~\cite{partridge_molecular_2005,wang_photoexcitation_2021}) which is $\SI{851}{MHz}$ blue detuned from the center of the $D_2$ line. Since we are not able to calculate the photoassociation absorption rate \textit{ab initio} we use the dissipation power as the main experimental variable. We confirm that atoms are lost in pairs by varying the spin polarization of the gas, finding that atom losses are negligible in a fully spin polarized cloud. Moreover, we find that the associated molecular state leaves the trap without additional interactions with the cloud, namely without significant heating in the reservoirs.

\subsection{Heating and atom loss}

\begin{figure}[t]
    \centering
    \includegraphics[width=85mm]{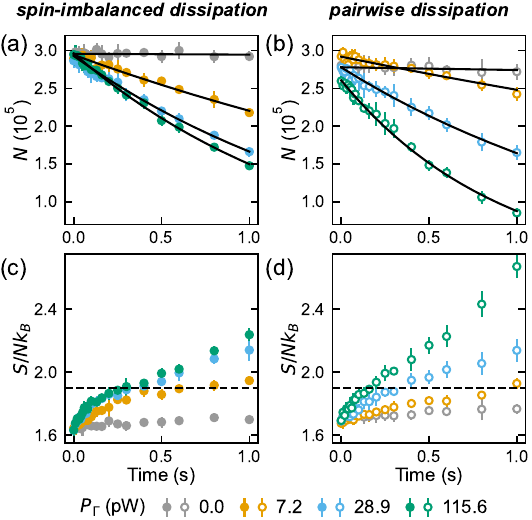}
    \caption{\textbf{Atom loss and heating with dissipation for the entropy transport experiment.} (a),(b) Total atom number as a function of time. The solid lines are fits to exponential decay. (c),(d) The entropy per particle increases over time mostly due to atom losses rather than heating, as temperatures remain almost the same (see Tab.~\ref{tab:heating}). The majority of data points is still below the superfluid transition $s_c\approx1.9$ (dashed line, see text) in our reservoirs. }
    \label{fig:figS_losseffect}
\end{figure}

\begin{figure}[t]
    \centering
    \includegraphics[width=85mm]{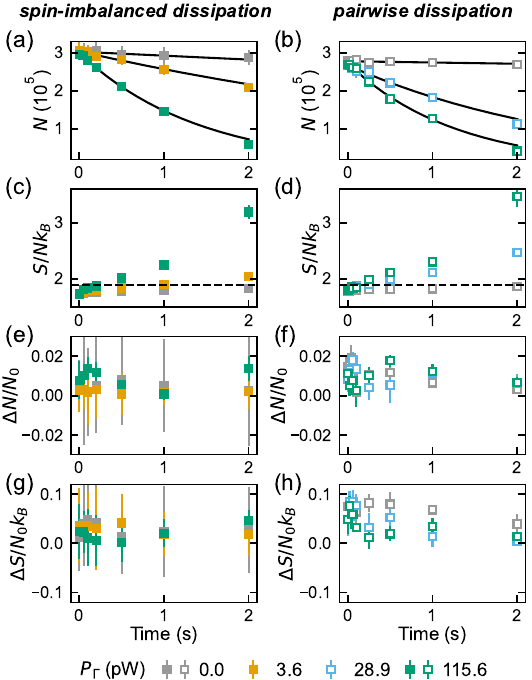}
    \caption{\textbf{Atom loss, heating, and imbalances for the spin transport experiment.} (a),(b) Total atom number as a function of time. The solid lines are fits to exponential decay. (c),(d) Entropy per particle. The dashed line indicates the superfluid transition $s_c\approx1.9$ in the reservoirs. (e)-(h) Particle and entropy imbalances are prepared close to zero and barelly evolve over time.}
    \label{fig:figS_spin}
\end{figure}

The observed initial and final thermodynamic properties of the reservoirs are summarized in Tab.~\ref{tab:heating} for all experiments discussed in the main text. The dynamics of atom number $N$ and entropy per particle $S/N$ in the entropy transport experiment are displayed in Fig.~\ref{fig:figS_losseffect}. Those for the spin transport experiment are shown in Figs.~\ref{fig:figS_spin}(a)--(d). We find that the engineered dissipation does not lead to significant heating, $T(t=0)\approx T(t=\SI{1}{s})$. The observed reduction of degeneracy stems predominantly from the atom loss. Importantly, the majority of the data points is still below the critical entropy per particle $s_c/k_B\approx 1.9$~\cite{fabritius_irreversible_2024} in the slightly anharmonic trap during the transport experiment, which is slightly higher than the harmonic limit $s_c^\mathrm{ho}/k_B\approx 1.69$~\cite{ku_revealing_2012}. This suggests that heating the reservoirs beyond the critical point is not responsible for most of the observed changes in the transport dynamics. The observed enhanced diffusion is already significant for the smallest dissipation where the reservoirs stay superfluid throughout the full transport dynamics. Moreover, the low thermal conductance in the absence of dissipation has been observed previously~\cite{husmann_breakdown_2018} at a degeneracy of $q=\mu/k_B T=1.5(2)$ which is comparable to the lowest degeneracy measured here $q=1.2(3)$ (see Tab.~\ref{tab:heating}A). Thus even above the superfluid transition the strong attractive interactions and the local pairing gap~\cite{husmann_breakdown_2018,krinner_mapping_2016} (enhanced by the gate potential) prevent spin and thermal diffusion. 
The total local potential around the channel $V_\mathrm{tot}=V_g+\mu$ is weakly affected by dissipation since it is dominated by the gate potential $V_g\gg\mu$. This means that the number of occupied transport modes $n_\mathrm{m}$ is almost constant during transport. For $\nu_x=\SI{10.9}{kHz}$ the number of thermally occupied modes, calculated assuming Fermi-Dirac distribution of the reservoirs and the expected energy landscape of the channel (Methods of Ref.~\cite{fabritius_irreversible_2024}), reduces from $n_\mathrm{m}(t_\mathrm{i})\approx3.1$ to $n_\mathrm{m}(t_\mathrm{f})\approx2.7$, using the thermodynamic quantities in Tab.~\ref{tab:heating}A.

\begin{table}
    \centering
    \begin{tabular}{|c|c|c|}
    \hline
         \textbf{A}& $t_\mathrm{i}=\SI{0}{s}$ & $t_\mathrm{f}=\SI{1}{s}$ \\ \hline
         $S/N k_B$& $1.65(3)$& $2.6(3)$ \\ \hline
         $N$& $2.86(15)\times 10^5$& $1.0(3)\times 10^5$ \\ \hline
         $T$& $\SI{92(2)}{nK}$&  $\SI{99(2)}{nK}$ \\ \hline
         $\mu/k_B$& $\SI{235(5)}{nK}$& $\SI{113(30)}{nK}$ \\ \hline
    \hline
         \textbf{B}& $t_\mathrm{i}=\SI{0}{s}$ & $t_\mathrm{f}=\SI{1}{s}$ \\ \hline
         $S/N k_B$& $1.75(4)$& $2.27(5)$ \\ \hline
        $N$& $2.9(2)\times 10^5$&  $1.4(1)\times 10^5$ \\ \hline
         $T$& $\SI{97(2)}{nK}$&  $\SI{97(2)}{nK}$ \\ \hline
         $\mu/k_B$& $\SI{232(7)}{nK}$& $\SI{152(7)}{nK}$ \\ \hline
    \end{tabular}
    \caption{\textbf{Overview of initial and final thermodynamic quantities.} Table A shows the values measured for the entropy transport experiment (see Fig.~2 and Fig.~\ref{fig:figS_losseffect}). Table B shows the quantities for the spin transport experiment (see Fig.~3 and Fig.~\ref{fig:figS_spin}). For the latter, although the data are taken up to $t=\SI{2}{s}$, we show here properties at $t_\mathrm{f}=\SI{1}{s}$ for comparison with (A).
    The initial properties of the reservoirs at $t_\mathrm{i}$ are averages over all datasets in the corresponding experiment. The final properties at $t_\mathrm{f}$ are calculated for the datasets with the strongest dissipation $P_\Gamma=\SI{115.6}{pW}$, averaging the spin-imbalanced and pairwise dissipations. }
    \label{tab:heating}
\end{table}

\section{Fitting procedure}
\subsection{Atom number fit}

The overall atom loss rate is determined from a fit of the atom number evolution $N(t)$ shown in Figs.~\ref{fig:figS_losseffect}(a),(b) and Figs.~\ref{fig:figS_spin}(a),(b)
with an exponential decay function given by
\begin{equation}
    N_\mathrm{fit}(t) = N_0 e^{-\gamma_N t }\,.\label{eq:nfit}
\end{equation}
The initial value $N_0$ and the loss rate $\gamma_N$ are fitted using the least-squares method. The sum of squared residuals is calculated from the mean and weighted by the inverse of the standard deviation of each measured $N(t)$. The average loss current from each reservoir $I_\Gamma$ is then defined as $I_\Gamma(t)=-\dot{N}(t)/2=\gamma_N N(t)/2$ such that
\begin{equation}
    I_\Gamma(0)=\gamma_N N_0/2\,.\label{eq:loss-current}
\end{equation}

\subsection{Magnetization imbalance fit}
The spin conductance is determined from a fit of the magnetization imbalance evolution $\Delta M(t)$
with an exponential decay function given by
\begin{equation}
    \Delta M_\mathrm{fit}(t) = \Delta M_0 e^{- t/\tau_\sigma}\,.\label{eq:dmfit}
\end{equation}
The initial imbalance $\Delta M_0$ and the timescale of spin transport $\tau_\sigma$ are fitted using the least-squares method. The spin conductance $G_\sigma$ in the spin-degenerate system is then given by 
\begin{equation}
    G_\sigma = \frac{\chi}{4\tau_\sigma},\label{eq:gsigma}
\end{equation}
where the spin susceptibility is defined as $\chi=2\pdv{(N_\uparrow-N_\downarrow)}{(\mu_\uparrow-\mu_\downarrow)}$. The spin susceptibility is calculated from the initial ($t=0$) atom number, temperature and magnetization. We did not find an appreciable difference when replacing the initial thermodynamic quantities by the average values over the full transport time. The spin conductance in the non-interacting, non-dissipative system for a ballistic 1D junction is $G_{\sigma,0} = 2 n_\mathrm{m}/h$.

\subsection{Phenomenological model fit}
The differential equations obtained from the phenomenological model are valid for the spin-degenerate system, where the reservoir response coefficients are calculated for the full-harmonic reservoirs (both spins in both reservoirs), for more detail see the Supplementary Material of Ref.~\cite{fabritius_irreversible_2024}.
The equations we use for the fit of the dynamics describing particle and entropy transport are given by
\begin{align}
    -\frac{1}{2}\dv{\Delta N(t)}{t} &= I_{\mathrm{exc}} \tanh{\left(\frac{\Delta \mu+\alpha_c \Delta T}{\sigma}\right)}\,,\\
    -\frac{1}{2}\dv{\Delta S(t)}{t} &= -\frac{\alpha_c}{2} \dv{\Delta N(t)}{t}  + G_T \frac{\Delta T}{T}\,.
\end{align}
The reservoir response is given by
\begin{align}
    \Delta\mu &= \frac{(\ell_r+\alpha_r^2)\Delta N  - \alpha_r \Delta S} {\ell_r \kappa/2}\,,\\
    \Delta T &=\frac{-\alpha_r \Delta N + \Delta S} {\ell_r \kappa/2}\,,
\end{align}
where $\kappa=(\partial N/\partial\mu)_T$ is the compressibility, $\alpha_r=(1/\kappa)(\partial S/\partial \mu)_T$ is the dilatation coefficient, and $\ell_r=C_N/T\kappa$, with $C_N=T(\partial S/\partial T)_N$, is the ``Lorenz number'' of the reservoirs \cite{grenier_thermoelectric_2016}.
The equations are integrated numerically in order to 
fit the measured evolution of entropy and atom number imbalances. We reduce the fluctuations in the data of $\Delta N(t)$ and $\Delta S(t)$ by multiplying the relative imbalances $\Delta N(t) / N (t)$ and $\Delta S(t) / N (t)$ from each experimental realization with the fitted atom number $N_\mathrm{fit}(t)$ (Eq.~\ref{eq:nfit}).
For datasets with strong dissipation where the system relaxes to equilibrium [the two highest $P_\Gamma$ in Figs.~2(b),(d), and the highest $P_\Gamma$ in Figs.~2(c),(e)], we subtract small offsets in $\Delta N/N_0$ and $\Delta S/N_0k_B$ using the data points at the longest $t$, in order to account for possible misalignment of the experiment and to improve the fit. 

The reservoir response coefficients are not fitted directly, but the coefficients computed via the equation of state (EoS) are multiplied by a scaling factor which is fitted to account for fluctuations in atom number and temperature between datasets:
\begin{equation}
    \alpha_r = \alpha_s \alpha_\mathrm{EoS}\,,\quad
    \ell_r = \ell_s \ell_\mathrm{EoS}\,,\nonumber    
\end{equation}
where the indices stand for: $s$ -- scaling factor, EoS -- computed from EoS. The reservoir coefficients are calculated using the initial atom number and entropy. The initial values $\Delta N(0)$ and $\Delta S(0)$ are fixed from the data.
Initial parameters for the thermal conductance $G_T$ are determined from a diffusive timescale $\tau_d$, obtained from an exponential fit starting at the turning point (maximum $\Delta N$) of the particle imbalance evolution [Figs.~2(b),(c)]. They are related, in the limit that the advective transport timescale is much shorter than the diffusive timescale, by
\begin{equation}
    G_T\approx \frac{ T\kappa(\ell_r+(\alpha_c-\alpha_r)^2) }{4\tau_d } \,.
    \label{eq:GT_taud}
\end{equation}

The fit parameters are listed in Table~\ref{tab:dissi-fit_parameters}. Only the Seebeck coefficient $\alpha_c$, the thermal conductance $G_T$ and the nonlinearity coefficient $\sigma$ are fitted individually for each dataset. The scaling factors of the Lorentz number $\ell_s$ and dilatation coefficient $\alpha_s$ of the reservoirs are fitted simultaneously for all datasets (varying the strength and type of dissipation, as well as transverse confinement $\nu_x$ of the channel). 
Note that both $\alpha_c$ and $\alpha_r$ could be affected by dissipation as we discussed in the main text. However, only $\alpha_c-\alpha_r$ is relevant for the dynamics so we made a choice to fix $\alpha_r$ among datasets in the fit to limit the degrees of freedom. The fit result gives $\ell_r/k_B^2=0.028(1)$ and $\alpha_r/k_B=1.19(4)$.

\begin{figure}
\centering
\includegraphics[width=85mm]{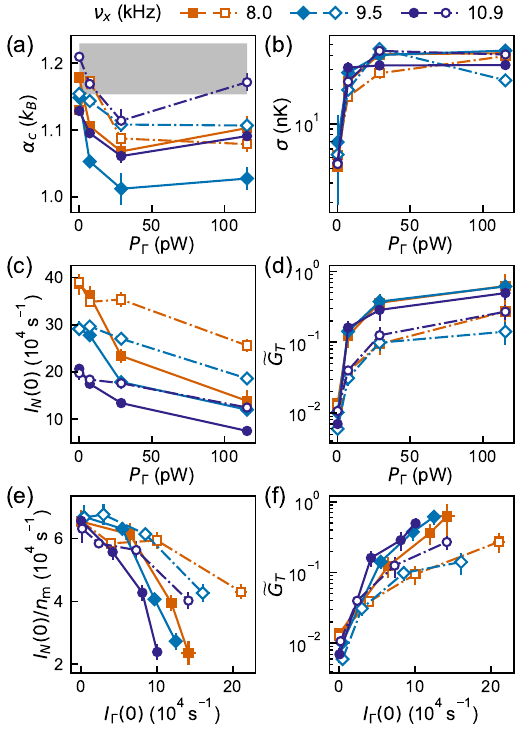}
\caption{\textbf{Fitted transport coefficients for extended datasets in the entropy transport experiment.} Filled (open) symbols again represent spin-imbalanced (pairwise) dissipation. The fitted Seebeck coefficient (a) as a function of dissipation beam power is slightly decreasing for weak dissipation. The gray bar represents the average fitted $\alpha_r=1.19(4)$, showing that $\alpha_r-\alpha_c$ increases with weak dissipation. (b) The nonlinearity coefficient $\sigma$ increases sharply with dissipation, indicating a more linear transport response. (c),(d) Initial advective particle current $I_N(0)$ and normalized thermal conductance $\widetilde{G_T}$ versus dissipation strength.
Errorbars are standard deviations from the least squares fit. Current and conductance of (c),(d) are replotted in (e),(e) versus the initial loss current on the horizontal axis. The initial currents (e) are also normalized by the number of transverse mode $n_\mathrm{m}$. (e),(f) serve to compare the two dissipation mechanisms, showing that pairwise dissipation generally has a weaker effect on transport for a given particle loss rate. }
\label{fig:figS_params}
\end{figure}

Moreover, the compressibility $\kappa$ is fixed to the computation from the EoS. The excess current $I_\mathrm{exc}$ is fixed from the initial current $I_N(0)$, which is determined from a linear fit to the initial particle imbalance evolution. The linear fit includes between $4$ and $10$ data points. Starting with 4 points the linear fit is repeated with increasing number of points until the reduced $\chi^2$ is minimized. The residual sum of squares for minimization is a concatenation of deviations in $\Delta N$ and $\Delta S$ of all datasets,
with weights given by the inverse averaged standard deviations of $\Delta N$ and $\Delta S$ respectively for each dataset.
In addition to the 8 datasets shown in Fig.~2 (4 datasets per dissipation type) $16$ additional datasets (8 datasets per transverse confinement) are fitted simultaneously (Fig.~\ref{fig:figS_params}). The auxiliary datasets were measured for confinement frequencies of $\nu_x=\SI{9.5(2)}{kHz}$ and $\nu_x=\SI{8.0(2)}{kHz}$, within the quasi-1D regime. All datasets can be well described with the phenomenological model, reaching a goodness of fit parameter of reduced $\chi^2$ of 0.65.

The excess current $I_\mathrm{exc}$, Seebeck coefficient $\alpha_c$, nonlinearity coefficient $\sigma$ and normalized thermal conductance $\widetilde{G}_T$ for all datasets are shown in Fig.~\ref{fig:figS_params}. We find that the results displayed in Fig.~4 are reproduced for various confinement frequencies and thus different number of occupied modes. 

\begin{table}[ht]
    \centering
    \begin{tabular}{|c|c|}
    \hline
    fitted for each dataset & $\alpha_c$, $G_T$,  $\sigma$  \\
    \hline
    shared among all datasets & $\ell_s$, $\alpha_s$ \\
    \hline
    fixed & $\kappa$, $I_\mathrm{exc}$ \\
    \hline
    \end{tabular}
    \caption{\textbf{Fit parameters.}}
    \label{tab:dissi-fit_parameters}
\end{table}

The obtained conductance is normalized with $G_{T,0}=2n_\mathrm{m}\pi^2 k_B^2 T/3 h$ in Fig.~4(c) and Fig.~\ref{fig:figS_params}(d) such that $\widetilde{G}_T$ is independent of temperature $T$. The number of occupied channel modes $n_\mathrm{m}\approx 5.9,\,4.4,\,3.1$ for $\nu_x=8.0,\,9.5,\,10.9$ respectively, is calculated assuming Fermi-Dirac distribution of the reservoirs and the expected energy landscape of the channel (see Methods of Ref.~\cite{fabritius_irreversible_2024}). 

The current $I_\mathrm{n0}$ for a non-interacting system shown in Fig.~4(b) (dotted horizontal line) 
is calculated assuming a chemical potential bias of $\Delta \mu = \SI{50}{nK}$, which corresponds to the largest $\Delta N$ observed, and the number of modes given by $n_\mathrm{m}$ such that $I_\mathrm{n0}=2n_\mathrm{m}\Delta \mu/h$. The initial currents displayed in Fig.~\ref{fig:figS_params}(c) differ mostly due to the different number of occupied modes. In (e) we normalize the current by $n_\mathrm{m}$, showing that the measured initial currents fall on top of each other without dissipation.

\section{Comparison of spin-imbalanced and pairwise dissipation}

Figures~\ref{fig:figS_params}(e),(f) replot the initial particle current and the normalized thermal conductance $\widetilde{G}_T$ versus the initial loss current. Plotting them as functions of the loss current---equivalent to an overall loss rate---offers one way to compare the effects of spin-imbalanced (filled symbols) and pairwise dissipation (open symbols). We find that at an equivalent loss current, pairwise losses suppress the particle current less than spin-imbalanced losses. Likewise, pairwise losses enhance the thermal conductance less than spin-imbalanced losses. This shows that the nature of the dissipation plays a role in slowing down the particle transport and in enabling entropy diffusion. 
As mentioned in the main text, 
a possible mechanism is that particle current depends on the locally suppressed superfluid order and the timescale with which the order can be re-established via the reservoirs. In this picture pairwise losses, which act on the superfluid order parameter~\cite{partridge_molecular_2005}, can be directly replenished by the superfluid reservoirs. In contrast, the spin-imbalanced loss creates some local unpaired quasi-particle excitation above the superfluid gap with excess energy that could be more destructive to the the superfluid order.

\section{Discussion on the uncertainties of the fitted conductances}

\subsection{Thermal conductance}

Since $G_T$ results from a fit to the phenomenological model, it might be dependent on the fitted reservoir response coefficients. 
In particular, as noted previously~\cite{fabritius_irreversible_2024} the fitted $\ell_r$ strongly deviates from the theoretical value from the 3D EoS for harmonically trapped reservoirs, indicating that these response functions depend on details of the potential landscape. 

It can be shown that within our phenomenological model, in the limit of small $G_T$ (non-dissipative case), $\ell_r$, $\alpha_c$ and $\alpha_r$ are constrained by the maximum $\Delta N$ reached after preparing an initial $\Delta S_0>0$ by
\begin{equation}
    \Delta N_\mathrm{max}\approx\frac{\alpha_r-\alpha_c}{\ell_r+(\alpha_r-\alpha_c)^2}\Delta S_0.
    \label{eq:DeltaNmax}
\end{equation}

\begin{figure}[t]
\centering
\includegraphics[width=85mm]{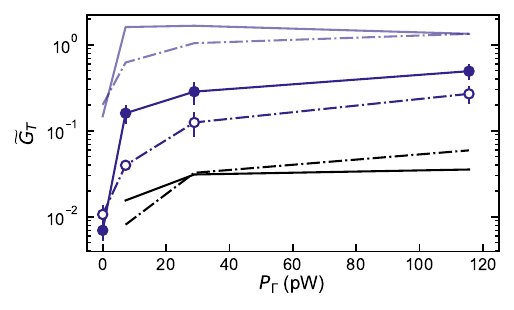}
\caption{\textbf{Limits of fitted thermal conductances.} 
Data points are the same as in Fig.~4(c). Filled (open) symbols correspond to spin-imbalanced (pairwise) dissipation. Lines in lighter navy correspond to the estimated upper bound within the phenomenological model given by Eq.~\ref{eq:GT_upperbound}. Black lines correspond to the apparent thermal conductances assuming no real transport but only particle losses from the reservoirs given the measured particle loss rate.}
\label{fig:figS_GT_bounds}
\end{figure}

In this limit, $\ell_r$ has a maximum value of $(\Delta S_0/\Delta N_\mathrm{max})^2/4$ and $\alpha_r-\alpha_c \in (0,\Delta S_0/\Delta N_\mathrm{max})$. The bounds of $\alpha_r-\alpha_c $ are approached when $\ell_r\rightarrow 0$. From Eq.~\ref{eq:DeltaNmax} alone, $\alpha_r-\alpha_c$ could be either below or above $\Delta S_0/(2\Delta N_\mathrm{max})$. However, we can put an upper bound for $\alpha_r< S/N$ based on physical arguments. $\alpha_r=(\partial S/\partial N)_T$ has the physical meaning of the entropy per particle when removing atoms from the reservoir through the channel while maintaining the temperature. At the center of the trap where the channel is connected, the entropy per particle is smaller than the average entropy per particle of the whole reservoir. This generally excludes the upper branch for $\alpha_r-\alpha_c >\Delta S_0/(2\Delta N_\mathrm{max})$. Our fits do favor small $\alpha_r-\alpha_c$ in the regime where $\alpha_r-\alpha_c\approx (\Delta N_\mathrm{max}/\Delta S_0)\ell_r$. 
Therefore, we can use $\alpha_r< S/N$ where $S/N$ is from the non-dissipative data, $\alpha_c$ from the fitted entropy per particle of the advective mode in the absence of dissipation, and the independently fitted diffusive timescale $\tau_d$ (exponential decay after the initial response) to obtain an upper bound for $G_T$ using Eq.~\ref{eq:GT_taud} and Eq.~\ref{eq:DeltaNmax},
\begin{equation}
    G_T\approx \frac{T\kappa (\alpha_r-\alpha_c)\Delta S_0}{4\tau_d \Delta N_\mathrm{max}}
    < \frac{T\kappa (S_0/N_0-\alpha_c)\Delta S_0}{4\tau_d \Delta N_\mathrm{max}} \,.
    \label{eq:GT_upperbound}
\end{equation}
In Fig.~\ref{fig:figS_GT_bounds} we plot the same $\widetilde{G}_T$ as in Fig.~4, with the corresponding upper bounds from Eq.~\ref{eq:GT_upperbound} shown in the same lineshape with a lighter color. These bounds are also very close to one, the non-interacting limit without dissipation.

On the other hand, we plot the equivalent thermal conductance from an apparent transport due to pure losses from both reservoirs (black lines in the corresponding lineshapes below the data). That is, without any real transport, $\Delta N$ will decay exponentially with rate $\gamma_N$ if each reservoir decays exponentially at rate $\gamma_N$. However, this scenario would give a much smaller thermal conductance than the observed diffusion so the observation cannot be attributed to losses.

\subsection{Spin conductance}
Here we discuss the uncertainty in the obtained spin conductances, particularly concerning the apparent spin transport induced by atom losses. 

Given that the two dissipation mechanisms lead to very similar spin conductances with respect to the loss current [Fig.~4(c), inset], we explain the idea using pairwise dissipation for simplicity. 
We can consider two extreme scenarios: 
1)
If the pairwise dissipation can only remove a pair with initial correlation, i.e., both atoms are arriving from the same reservoir, then the dissipation cannot change the magnetization $M_i$ in either reservoir. Thus the losses do not directly induce an apparent spin current. 
2)
If the pairwise dissipation \emph{always} removes a pair with constituents coming out of different reservoirs [in the picture of Fig.~3(a), always photoassociating a $\ket{\uparrow}$ atom from the left reservoir with a $\ket{\downarrow}$ atom from the right reservoir], dissipation alone leads to apparent spin current. Effectively, each spin exhibits a loss-induced apparent decay of $\Delta N_{\uparrow(\downarrow)}(t)=\Delta N_{\uparrow(\downarrow)}(0)e^{-\gamma_Nt}$. This corresponds to $\Delta M(t)=\Delta M_0e^{-\gamma_Nt}$. In fact, $\gamma_N$ is very close to $1/\tau_\sigma$, the observed decay rate of $\Delta M$.

However, considering that the channels length is considerably larger than the dissipation beam size, the first scenario appears physically much more plausible than the second scenario. Thus, we believe that the observed spin diffusion is real transport rather than purely arising from losses.

\end{document}